\documentclass[3p,times]{elsarticle}

\usepackage{ecrc}

\volume{00}

\firstpage{1}

\journalname{Nuclear Physics A}

\runauth{}

\jid{npa}

\jnltitlelogo{Nuclear Physics A}

\usepackage{amssymb}

\usepackage[figuresright]{rotating}

\begin{document}
\def\Journal#1#2#3#4{{#1} {\bf #2}, #3 (#4)}

\def\NCA{Nuovo Cimento}
\def\NIM{Nucl. Instr. Meth.}
\def\NIMA{{Nucl. Instr. Meth.} A}
\def\NPB{{Nucl. Phys.} B}
\def\NPA{{Nucl. Phys.} A}
\def\PLB{{Phys. Lett.}  B}
\def\PRL{Phys. Rev. Lett.}
\def\PRC{{Phys. Rev.} C}
\def\PRD{{Phys. Rev.} D}
\def\ZPC{{Z. Phys.} C}
\def\JPG{{J. Phys.} G}
\def\CPC{Comput. Phys. Commun.}
\def\EPJ{{Eur. Phys. J.} C}
\def\PR{Phys. Rept.}
\def\JHEP{JHEP}

\begin{frontmatter}



\dochead{}

\title{Di-lepton production in p+p collisions at $\sqrt{s_{_{NN}}} = 200$ GeV from STAR}


\author{Lijuan Ruan (for the STAR Collaboration)}

\address{Physics Department, Brookhaven National laboratory, Upton NY 11973}

\begin{abstract}
The di-electron analysis for 200 GeV p+p collisions is presented
in this article. The cocktail simulations of di-eletrons from
light flavor meson decays and heavy flavor decays are reported and
compared with data. The perspectives for di-lepton measurements in
Au+Au collisions are discussed.
\end{abstract}


\end{frontmatter}


\section{Introduction}
Ultra-relativistic heavy ion collisions provide a unique
environment to study properties of strongly interacting matter at
high temperature and high energy density~\cite{starwhitepaper}.
One of the crucial probes of this strongly interacting matter are
di-lepton measurements in the low and intermediate mass region.
Di-leptons are not affected by the strong interaction once
produced, therefore they can probe the whole evolution of the
collision. The di-lepton spectra in the intermediate mass range
($1.1\!<M_{ll}\!<\!3.0$ GeV/$c^{2}$) are directly related to
thermal radiation of the QGP~\cite{dilepton,dileptonII}. In the
low mass range ($M_{ll}\!<\!1.1$ GeV/$c^{2}$), we can study vector
meson in-medium properties through their di-lepton decays, where
any modifications observed may relate to the possibility of chiral
symmetry restoration. Measurements in p+p collisions provide a
crucial reference for the corresponding measurements in heavy ion
collisions. At STAR, the newly installed Time-of-Flight detector
(TOF) offers high acceptance and efficiency~\cite{startof}. The
TOF, combined with measurements of ionization energy loss (dE/dx)
from the Time Projection Chamber (TPC)~\cite{startpc}, enables
electron identification (eID) with high purity from low to
intermediate $p_T$~\cite{starelectron}. In this article we present
the di-lepton continuum from 200 GeV p+p collisions carried out in
2009. We utilized eID from the TOF and TPC. The cocktail
simulations are presented and compared to the data. Finally,
future capabilities for di-lepton measurements at STAR in Au+Au
collisions are discussed.

\section{Data Analysis and Results}
At STAR~\cite{star}, there are three detectors which are used for
electron identification at mid-rapidity: the TPC, the TOF, and the
Barrel Electron-magnetic Calorimeter (BEMC)~\cite{starbemc}. The
TPC is the main tracking detector at STAR, measuring momenta and
path-lengths of particles created in the collisions. The dE/dx is
used for particle identification~\cite{bichsel,pidpp08}. The full
TOF system was installed in STAR in year 2009 and extends
$\pi$($K$) identification up to 1.6 GeV/$c$ and $p(\bar{p})$ up to
3 GeV/$c$~\cite{pidNIMA,tofPID}. By combining the velocity
($\beta$) information from the TOF and the dE/dx from the TPC,
electrons can be clearly identified from low to intermediate
$p_T$. In addition, the EMC is used for triggering high energy
photons and electrons based on the energy deposited in the
detector.

In Run 9 p+p collisions, the TOF azimuthal acceptance is 72\% at
$|\eta|\!<\!0.9$. In the di-electron analysis, we used 107 million
minimum-bias p+p events at $\sqrt{s_{_{NN}}} = 200$ GeV. The
collision vertex was required to be within 50 cm from TPC center
along the beam line. We selected electron candidates with a 99\%
purity by applying velocity and dE/dx cuts on tracks with
$p_T\!>\!0.2$ GeV/$c$. The velocity was required to be in the
range of $|1/\beta-1|\!<\!0.03$. The dE/dx cut is $p_T$ dependent
to ensure electrons have a 99\% purity in each $p_T$ bin. The
$e^{+}$ and $e^{-}$ pairs from the same events were combined to
reconstruct the invariant mass distributions ($M_{ee}$) marked as
unlike-sign distributions which contain both signal and
background. Two methods were used for background reconstruction,
like-sign and mixed-event techniques. In the like-sign technique,
the electron pairs with the same charge sign were reconstructed
from the same events. In the mixed-event technique used in this
analysis, unlike-sign pairs were reconstructed from different
events. In order to ensure the mixed events have similar
acceptances, we only mixed events which had collision vertices
within 5 cm of each other in the beam line direction. The
background pairs were formed from two tracks in different events.
\renewcommand{\floatpagefraction}{0.75}
\begin{figure}[htbp]
\begin{center}
\includegraphics[keepaspectratio,width=0.49\textwidth]{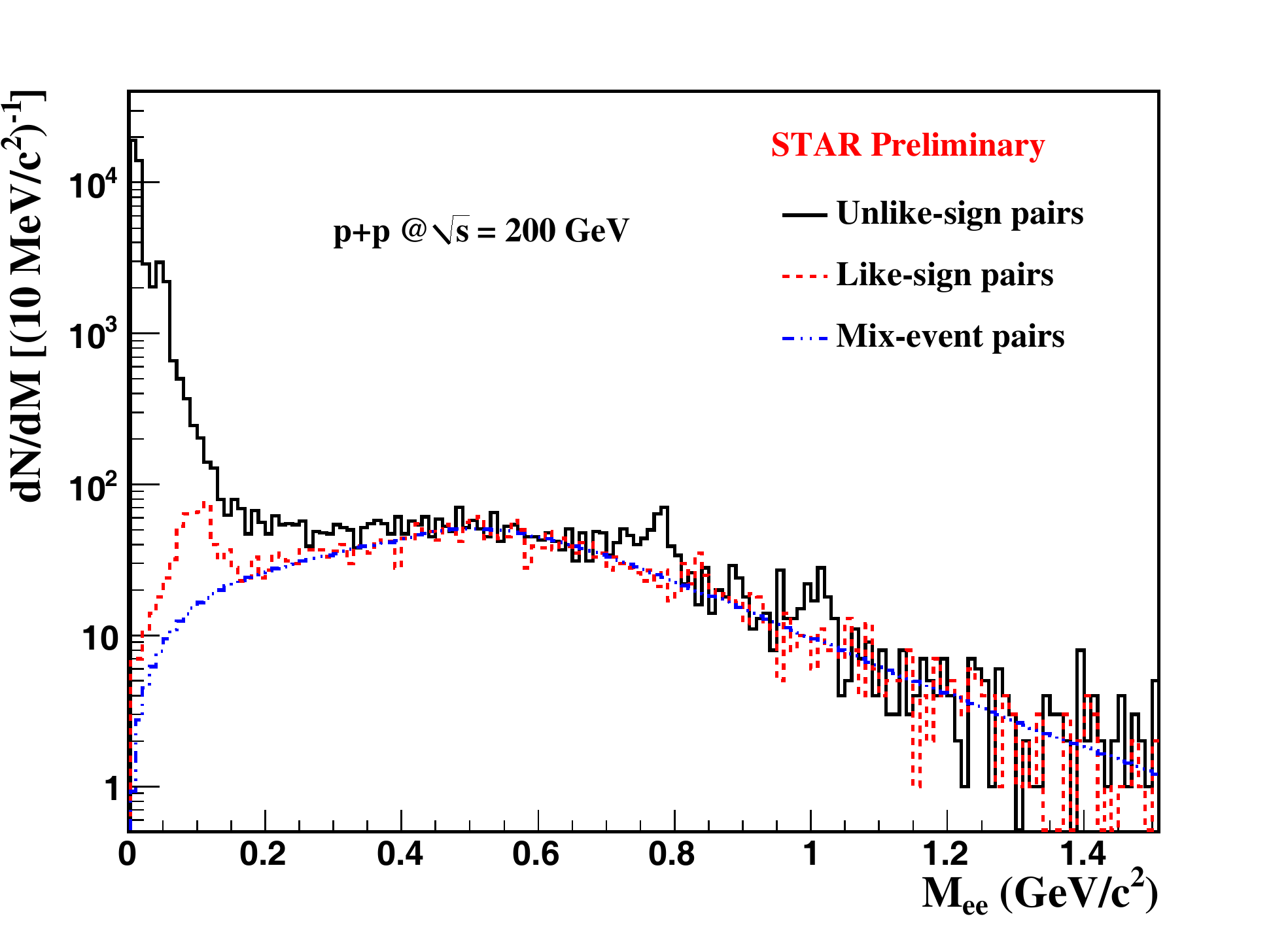}
\includegraphics[keepaspectratio,width=0.49\textwidth]{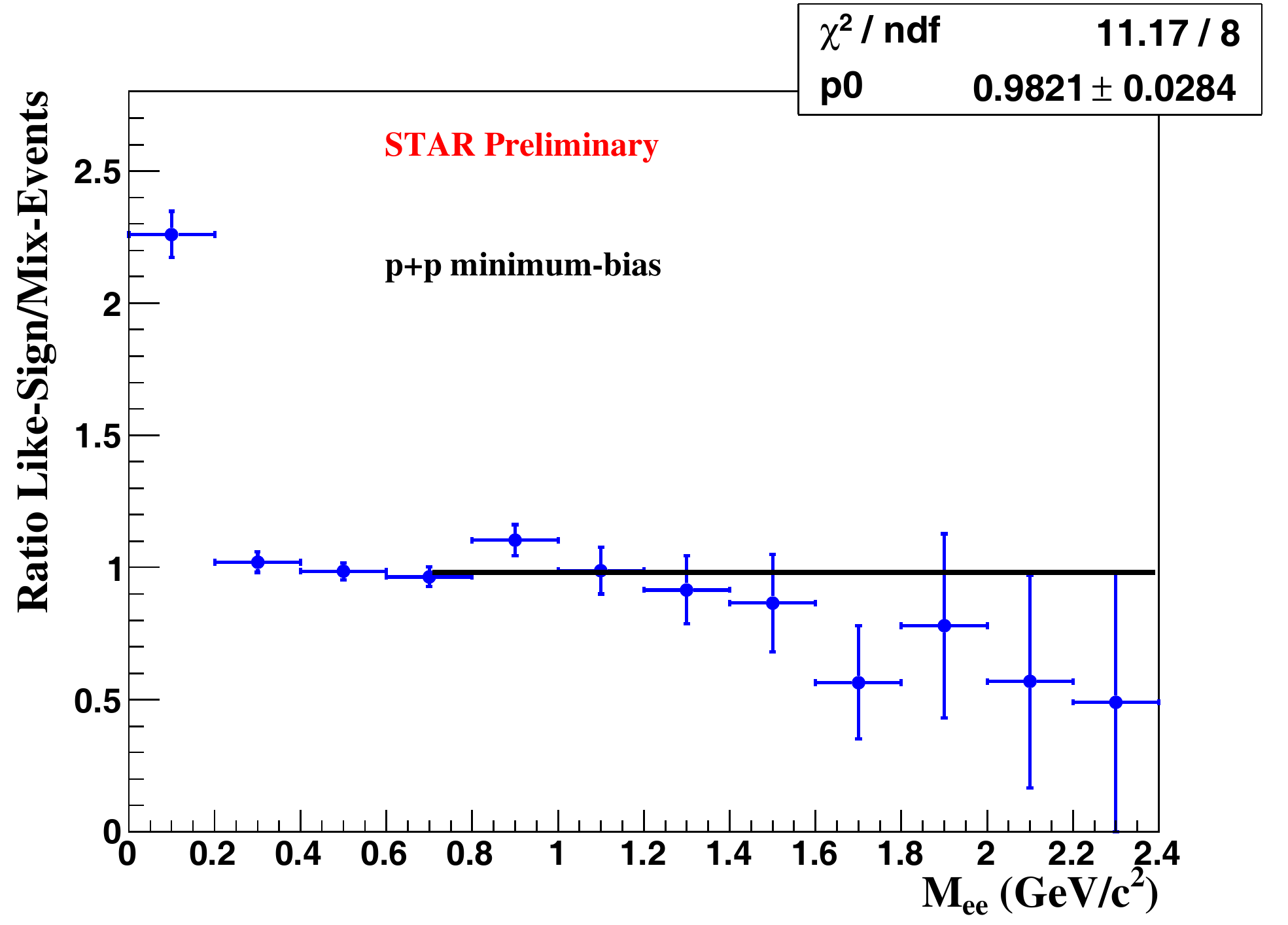}
\includegraphics[keepaspectratio,width=0.49\textwidth]{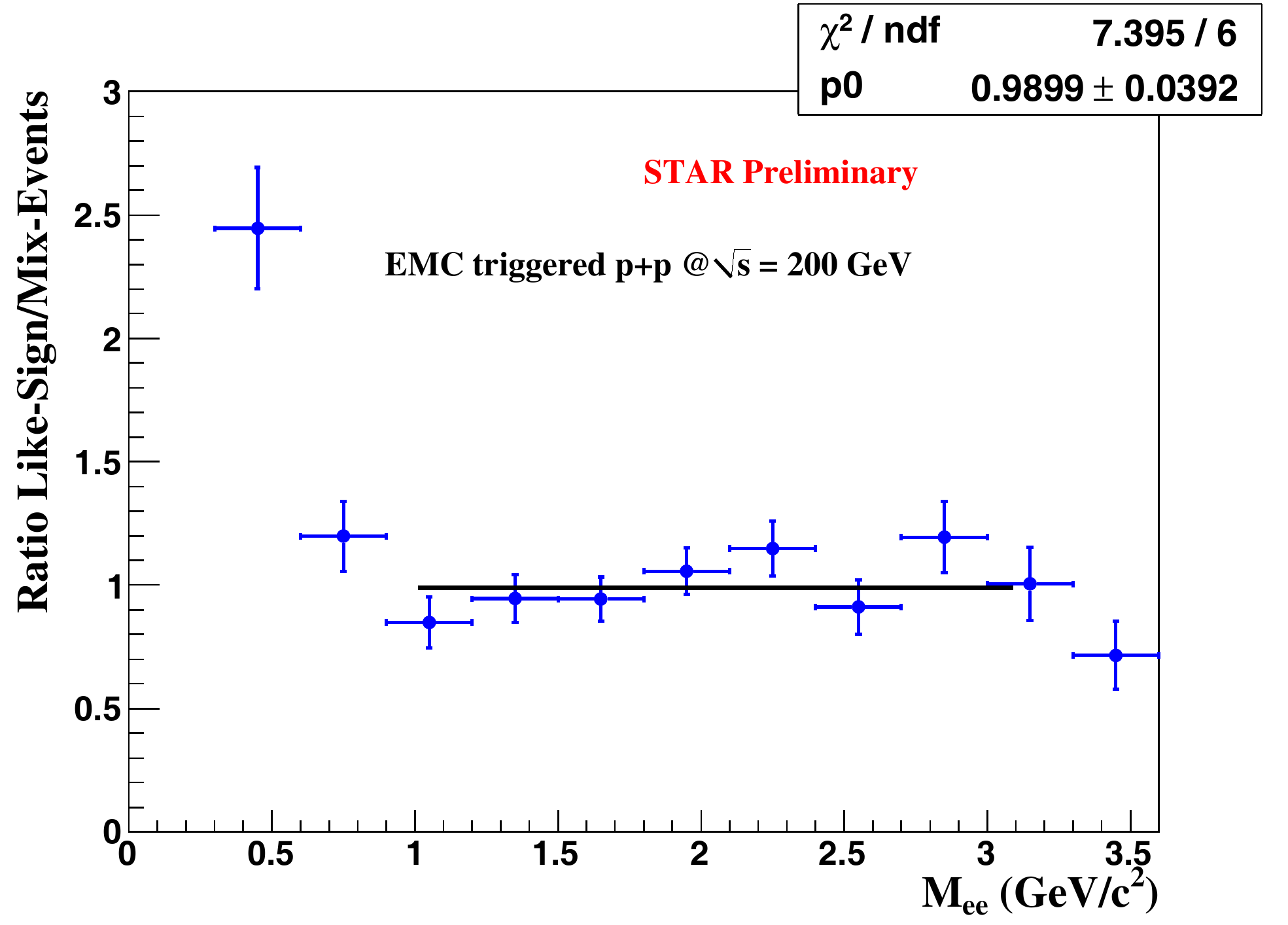}
\hspace*{-2cm}\begin{minipage}[t]{0.90\textwidth} \vspace*{-7mm}
\caption{(top-left panel) The electron pair invariant mass
distributions for unlike-sign pairs, like-sign, and mixed-event
background in minimum-bias p+p collisions. (top-right panel) The
ratio of like-sign over mixed-event distributions in minimum-bias
p+p collisions. (bottom panel) The ratio of like-sign over
mixed-event distributions in EMC triggered p+p collisions. }
\label{Fig:1}
\end{minipage}
\end{center}
\end{figure}

\subsection{Background subtraction}
Figure 1 (top-left panel) shows the invariant mass distribution
for unlike-sign pairs, like-sign, and mixed-event background. The
mixed-event distribution was normalized by a constant to the
like-sign distribution in the mass region of 0.4-1.5 GeV/$c^{2}$.
In the low mass region, there is correlated cross pair background,
which comes from double Dalitz decays, Dalitz decays followed by a
conversion of the decay photon, or two-photon decays followed by
conversion of both photons. This background is included in the
like-sign distribution but not in the mixed-event background. At
$M_{ee}\!<\!0.7$ GeV/$c^{2}$, the like-sign distribution was used
for background subtraction. At higher mass, we compared the shape
of like-sign and mixed-event distributions and found they matched
reasonably well, as shown in the ratio plot in Fig. 1 (top-right
panel). A constant was used to fit the ratio of like-sign over
mixed-event distributions and the $\chi^{2}/NDF$ is about
1~\cite{phenixppdilepton}. We did the same exercise using EMC
triggered events for di-electron analysis, in which one electron
is required to trigger the EMC and has a transverse momentum
greater than 2 GeV/$c$, and the $p_T$ of the other electron is
greater than 0.2 GeV/$c$. We found that with a significantly
higher transverse momentum for the pair, the shape of like-sign
over mixed-event distributions matched in the mass region of 1-3
GeV/$c^{2}$, as shown in the ratio plot in Fig. 1 (bottom panel).
This indicates that the jet contribution within the STAR
acceptance does not significantly alter the combinatorial
background shape in the mixed-event method and is negligible given
our precision. PYTHIA simulations with jets are on-going. For
minimum-bias events, at $M_{ee}\!>\!0.7$ GeV/$c^{2}$, we
subtracted the normalized mixed-event background. The di-electron
continuum after background subtraction is shown in Fig. 2 (right
panel). The errors shown are statistical only. The systematic
uncertainties on the normalization and background subtraction are
still under study. In the future, more simulation will be done to
further study background subtraction at $M_{ee}\!>\!0.7$
GeV/$c^{2}$.

\begin{figure}[htbp]
\begin{center}
\includegraphics[keepaspectratio,width=0.6\textwidth]{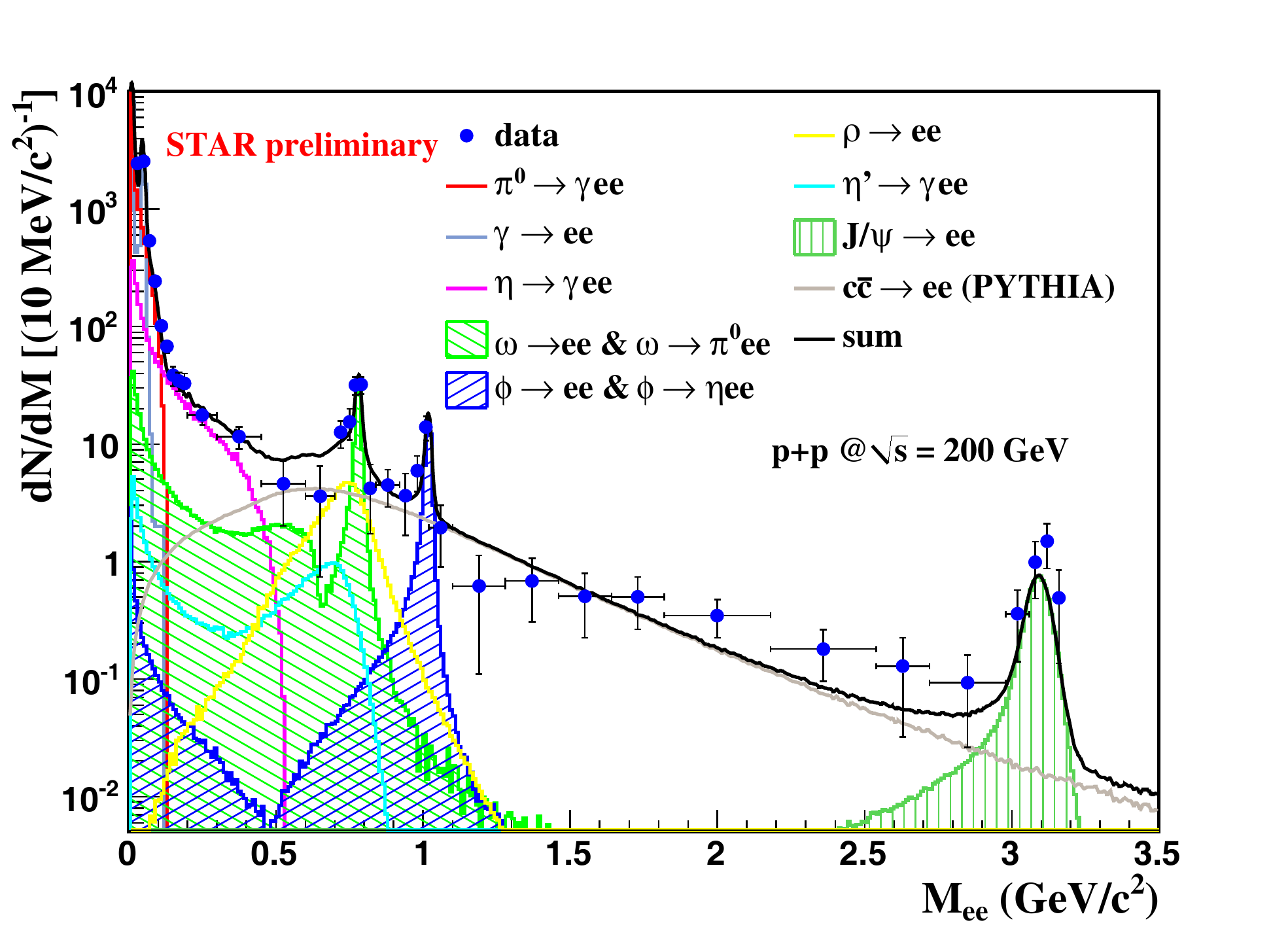}
\hspace*{-2cm}\begin{minipage}[t]{0.90\textwidth} \vspace*{-7mm}
\caption{The comparison for di-electron continuum between data and
simulation in 200 GeV minimum-bias p+p collisions. The di-electron
continuum from simulations with different source contributions are
also shown.} \label{Fig:2}
\end{minipage}
\end{center}
\end{figure}

\subsection{Cocktail simulation}
The di-electron signals may come from light flavor hadron decays
and heavy flavor hadron decays, for example, $\pi^{0}$, $\eta$,
and $\eta^{\prime}$ Dalitz decays: $\pi^{0}\rightarrow \gamma
e^{+}e^{-}$, $\eta \rightarrow \gamma e^{+}e^{-}$, and
$\eta^{'}\rightarrow \gamma e^{+}e^{-}$; vector meson decays:
$\omega \rightarrow \pi^{0} e^{+}e^{-}$, $\omega \rightarrow
e^{+}e^{-}$, $\rho^{0} \rightarrow e^{+}e^{-}$, $\phi \rightarrow
\eta e^{+}e^{-}$, $\phi \rightarrow e^{+}e^{-}$, and $J/\psi
\rightarrow e^{+}e^{-}$; heavy flavor decays: $c\bar{c}
\rightarrow e^{+}e^{-}$ and $b\bar{b} \rightarrow e^{+}e^{-}$; and
Drell-Yan contributions. We fit the invariant yields of measured
mesons with the Tsallis blast-wave functions~\cite{Tsallis}, used
the Tsallis functions as our inputs, and decayed them into
di-electrons with GEANT using STAR year 2009 geometry. The
electron candidates were reconstructed as tracks. We applied the
same cuts for the electron tracks as in the data and generated the
$e^{+}e^{-}$ cocktails from different contribution sources for the
same number of events and rapidity ranges as used in the data
analysis. The gamma conversion $\gamma\rightarrow e^{+}e^{-}$ was
not rejected and its contribution to di-electron continuum was
simulated with the same method. Compared to what was presented in
the conference, the Dalitz decays of $\omega \rightarrow \pi^{0}
e^{+}e^{-}$ and $\eta^{'}\rightarrow \gamma e^{+}e^{-}$ were
updated using the correct Kroll-Wada expression~\cite{krollwada};
and the $\rho^{0} \rightarrow e^{+}e^{-}$ line shape was updated
with the Boltzmann phase space
factor~\cite{rholineshape,starrho0pp}. The total contribution from
the simulation is shown as the black solid curve on Fig. 2 and is
found to be consistent with data. It is also consistent with what
was shown in the conference. The invariant yield of $\pi^{0}$ is
taken as the average of $\pi^{+}$ and
$\pi^{-}$~\cite{tofPID,ppdAuPID}. The yields of
$\phi$~\cite{starphipp} and $\rho^{0}$~\cite{starrho0pp} are from
STAR while $\eta$~\cite{phenixetapp},
$\omega$~\cite{phenixomegapp} and $J/\psi$~\cite{phenixjpsipp} are
from PHENIX. In this simulation, the $c\bar{c}$ cross section was
an input and constrained by the measurement from
STAR~\cite{starelectron,weihardprobes2010}. In the intermediate
mass region, di-electron continuum is dominated by the $c\bar{c}$
contribution. The $\chi^{2}/NDF$ of the comparison between data
and simulation is 36/30 in the mass region of 0.1-3.2 GeV/$c^{2}$.
The uncertainties on the cocktail including the decay form
factors, the measured cross section for each hadron, and the
efficiency uncertainties are under-study. The current precision of
STAR di-electron data presented here does not allow us to
distinguish STAR's measured charm cross section from
PHENIX's~\cite{starelectron,phenixelectron}. With D meson
reconstruction and non-photonic electron measurements at low $p_T$
from the TPC and TOF from year 2009 p+p collisions, the charm
cross section will be measured with a better precision at STAR and
can be used to further constrain the charm correlation
contribution to the di-electron continuum measurements.

\subsection{Future measurements in Au+Au collisions}

In year 2010, STAR has taken a few hundred million minimum-bias
events in 200, 62, and 39 GeV Au+Au collisions with full TOF
azimuthal coverage and low conversion material budget, which will
enable us to study the following physics topics: 1) di-electron
enhancement in the low mass region~\cite{phenixdielectron}; 2)
in-medium modifications of vector meson decays; 3) virtual
photons~\cite{phenixdielectronII}; 4) $c\bar{c}$ medium
modifications; and 5) possible thermal radiation in the
intermediate mass region. With the current data sets, it will be
difficult to measure 4) or 5) since they are coupled to each other
and one is the other's background for the physics case. So far at
RHIC, there is no clear answer about thermal radiation in the
intermediate mass region. The future detector upgrade with the
Heavy Flavor Tracker at STAR will provide precise charm cross
section measurements~\cite{hft}, however the measurements of
$c\bar{c}$ correlations will still be challenging if not
impossible. An independent approach is proposed with the proposed
Muon Telescope Detector~\cite{starmtdproposal}, $\mu-e$
correlations, to measure the contribution from heavy flavor
correlations to the di-electron or di-muon continuum. This will
make it possible to access the thermal radiation in the
intermediate mass region.

\subsection{Summary}
In summary, the di-electron continuum is measured in 200 GeV p+p
collisions at STAR. The cocktail simulations are consistent with
the data in 200 GeV p+p collisions, providing a reference for the
future Au+Au study. The newly installed TOF system enables this
study and will also provide interesting physics outputs with the
data taken in year 2010 in Au+Au collisions at different beam
energies.








\end{document}